\newcommand{\tmem}[1]{{\em #1\/}}
\newcommand{\mathd}{\mathrm{d}}
\newcommand{\tmmathbf}[1]{{\mathbf #1}}
\newcommand{\tmop}[1]{\operatorname{#1}}
\newcommand{\op}[1]{#1}

\documentclass{elsart}

\usepackage{amssymb}

\begin{document}

\begin{frontmatter}



\title{Covariant Thermodynamics of an Object with Finite Volume}


\author{Tadas K. Nakamura}

\address{Fukui Prefectural University, 910-1195, Fukui, Japan}

\begin{abstract}
A covariant way to define the relativistic entropy of a finite object
has been proposed. The energy-momentum of an object with finite volume
is not a covariant physical entity because of the relativity of
simultaneity. A way to correctly handle this situation is introduced
and applied to the calculation of entropy. The result together with
van Kampen-Israel theory gives simple and self-consistent relativistic
thermodynamics.
\end{abstract}

\begin{keyword}
special relativity \sep thermodynamics \sep entropy

Special relativity \sep thermodynamics

\PACS  03.30.+p \sep 05.90.+m
\end{keyword}
\end{frontmatter}


Theory of relativistic thermodynamics has a long and controversial
history (see, e.g., \cite{stuart} and references therein).  The
controversy seems to have been settled more or less by the end of
1960s \cite{yuen}, however, papers are still being published to this
date (e.g., \cite{now}). Among a number of theories proposed, the one proposed
by van Kampen and later refined by Israel gives covariant definitions
to thermodynamical quantities. Van Kampen \cite{kampen} proposed to
treat the momentum of an object as a thermodynamical parameter.
Relativistic heat was defined in a covariant way to this end, however,
he states ``an equivalent, but slightly more streamlined this
formalism consists in eliminating the concept of heat altogether ...'' 
at the end of his paper. Israel \cite{israel} reformulated the problem
in this line, and obtained simple and straightforward covariant
thermodynamics.

The author of the present paper believes this van Kampen-Israel
theory is one of the best, if not {\tmem{the}} best, solution to the
problem basically. However, there still remains one problem that has
not been cleared: problem of three dimensional volumes. The purpose of
the present study is to make a small correction to the van
Kampen-Israel theory on this point and, hopefully, to complete the
fully covariant theory of relativistic thermodynamics.

It is known that the volume of an object viewed from distinct
inertial frames are different physical entities that are not connected
each other by a Lorentz transformation. Consequently total
energy-momentum of an object in one frame is not connected to that in
another frame, i.e., energy-momentum of an object with a finite volume is
not a covariant entity.  

Gamba \cite{gamba} wrote a paper about the confusion caused by this
fact; he states ``... physicists have made the same mistake. The
examples are so numerous that to review them all one should have to
write a book, not an article.''  He deplores the same misunderstanding
have not been eliminated from physics to the date of his paper in
1967, although the problem itself has been reported with the correct
answer as early as 1923 \cite{fermi}. He would deplore more to find
the same misunderstanding in papers to this date.

This misunderstanding causes an erroneous explanation of the
energy-momentum in a finite volume, which considers the effect of
container walls causes the difference of energy-momentum
definition. The paper by van Kampen \cite{kampen} seems to have fallen
this pitfall. Israel \cite{israel} were well aware of the problem,
however, he did not examine it in detail.

There have been papers warning this problem in connection with
relativistic thermodynamics (e.g., \cite{yuen}), however, the answer
that tells us how to handle the problem has not been explicitly
given. Our solution in the present study is to abandon the unique
energy-momentum of an object, and treat it as a function of four
dimensional ``direction'' of the volume. Consequently other
thermodynamical quantities become functions of the direction, however,
it is shown the entropy is a unique constant as long as there is no
entropy flux across the boundary of the object.

Let us briefly review van Kampen-Israel theory from our own view
point. Hereafter Einstein summation convention is enforced; summation
with Greek letters runs from 0 to 3, and roman letter is runs from 1
to 3. Zero-th component of represents time, and the speed of light is
scaled as unity ($c = 1$). A four vector is denoted by a bar (e.g.,
$\bar{a}$) and each component is represented by indexes ($a_{\mu}$ or
$a^{\mu}$).

We start with rewriting non-relativistic thermodynamics to eliminate the
concept of heat, and then extend it to relativity. In non-relativistic
thermodynamics, the entropy $S$ of an object is defined as
\begin{equation}
  \mathd S = \beta \mathd U - \beta \pi \mathd V~,
\end{equation}
where $\beta =$ inverse temperature ($= 1 / k_B T$), $U =$ thermal energy,
$\pi =$ pressure, and $V =$ volume, of the object. Let us neglect the second
term of the right hand side assuming $\mathd V = 0$ for a while, and
concentrate on the first term. The volume change will be considered later.

We rewrite $U$ with more basic physical quantities namely, the total
energy $E$, momentum $\tmmathbf{P}$ and mass $M$, as
\begin{equation}
  U = E - \frac{\tmmathbf{P}^2}{2 M}~.
\end{equation}
Then (1) with $\mathd V = 0$ is rewritten as
\begin{equation}
  \mathd S = \beta \left( \mathd E - \frac{\tmmathbf{P}}{M} \cdot \mathd
  \tmmathbf{P} \right)~.
\end{equation}
The most straightforward way to extend the above expression to relativity is
\begin{equation}
  \mathd E - \frac{\tmmathbf{P}}{M} \cdot \mathd \tmmathbf{P} \rightarrow
  \frac{1}{\sqrt{1 -\tmmathbf{v}^2}} ( \mathd E - v_i \mathd P^i ) = u_{\mu}
  \mathd P^{\mu}~,
\end{equation}
where $\tmmathbf{v}$ is the three dimensional velocity and $\bar{u}$ is four
velocity of the object. Then the relativistic entropy becomes
\begin{equation}
  \mathd S = \beta u_{\mu} \mathd P^{\mu} = \beta_{\mu} \mathd P^{\mu}~.
\end{equation}
Here $\beta_{\mu} = \beta u_{\mu}$ is so called four inverse
temperature. The above expression is identical to the equation (25) of
van Kampen \cite{kampen} with $\mathrm{d} A_{V \mu} = 0$.

The very basic definition of temperature is based on the fact:
when two equilibrium objects have heat exchange (random energy
exchange), heat flows from the higher temperature one to the lower
temperature one. Provided the entropy is suitably defined, this
statement is paraphrased as: ``{\tmem{heat flows spontaneously only
when the total entropy increases}}.''

Let us generalize the above statement to relativity. Heat is a form of
energy in non-relativistic thermodynamics, where the energy and
momentum are distinct quantities. In relativity, however, the energy
and momentum are components of one physical entity, energy-momentum
four vector namely, and thus cannot be treated
independently. Therefore we must treat the energy-momentum exchange
between the objects, not energy alone. Consequently, the inverse
temperature must have four components, $\beta_{\mu} = \beta u_{\mu}$,
corresponding to each component of energy-momentum four vector.

Suppose two objects (denoted by I and II) with different four inverse
temperature $\bar{\beta}_{\mathrm{I}}$ and $\bar{\beta}_{\tmop{II}}$
have random energy-momentum exchange.  Then the above statement
becomes: ``\tmem{energy-momentum transfer of $\mathd P$ from I to II
takes place spontaneously only when the total entropy increases},''
which means
\begin{equation}
  \mathd S = ( \beta_{\mathrm{I} \mu} - \beta_{\tmop{II} \mu} ) \mathd P^{\mu}
  > 0~.
\end{equation}
This formulation can treat not only heat conduction but also frictional
momentum transfer.

In the above review, we have treated $\bar P$ as a four vector subject
to Lorentz transformation. As mentioned earlier, however, this has one
problem in defining the total energy-momentum of the object. Suppose
an object with a finite extent, such as a gas in a container or a
solid body. Usually the volume of this object viewed from a reference
frame $\Sigma$ is defined as the three dimensional cross section of
its world tube at $t =\tmop{constant}$, where $t$ is the temporal
coordinate ($= x_0$) in $\Sigma$. Let us refer this volume as
$V$. Hereinafter the word ``volume'' means a three dimensional cross
section of the world tube in the four dimensional Minkowski space.

When we look at the same object from another inertial frame $\Sigma'$
that is moving with the velocity $\bar u$ relative to $\Sigma$, its
volume (we refer as $V'$ hereafter) is the cross section of $t' =
\op{\tmop{constant}}$ ($t'$: temporal coordinate in $\Sigma'$). These
two volumes $V$ and $V'$ are distinct physical entities: when we view
$V'$ from $S$, it is the cross section of $u_0 t - u_i x^i =
0$. Consequently physical quantities in $V$ and $V'$ are not the same
in general. 

M{\o}ller \cite{moller} has argued that the energy-momentum of a
finite object can be treated as four vector if the object is
isolated. This is true, however, there are cases in which we wish to
apply thermodynamics even when the object in interest is not
isolated. For example, an object in a heat bath with a constant
pressure is not isolated because of the momentum flux (= pressure),
but still thermodynamics should be valid. Therefore we need an
expression of energy-momentum that is available in the pressence of
fluxes.

To this end, let us introduce a way to handle the volume
mathematically. Given a unit time-like vector $u_{\mu}$, a three
dimensional flat surface is defined as a set of points that satisfies
$u_{\mu} x^{\mu} = 0$. The volume $V ( \bar{u} )$ is defined as the
intersection of this flat plane and the world tube of the object. Thus
$\bar u$ can be interpreted as the normal vector that defines the
direction of the three-dimensional volume in the four dimensional
space. Then volume can be represented by a four vector
\begin{equation}
  V_{\nu} ( \bar{u} ) = u_{\nu}  \frac{V_0}{u_0^{\lambda} u_{\lambda}}~,
\end{equation}
where $V_0$ is the amplitude of three dimensional volume of the object
in its rest frame, and $\bar{u}_0$ is the four velocity of the object,
in other words, the unit vector along the temporal axis of the
comoving frame. Usually the direction of the volume is taken to be
along the time axis of the reference frame, i.e., $\bar u =
(1,0,0,0)$, however, this can be any time-like unit vector in general.

Now that the volume of the object is the function of $\bar{u}$, then the
energy-momentum within this volume must depend on $\bar{u}$ as $\bar{P}_{} (
\bar{u} )$. The energy-momentum density tensor $T_{\mu}^{\nu}$ is supposed to
be constant within $\bar{V} ( \bar{u} )$ when the object is in the thermal
equilibrium, then we can write
\begin{equation}
  P^\nu ( \bar{u} ) = \int_{\bar{V} ( \bar{u} )} T^{\nu}_{\mu}
  \textrm{d} \sigma_{\mu} = \frac{V_0 T^{\nu}_{\mu} u^{\mu}}
  {u_0^\lambda u_{\lambda}}~.
\end{equation}

Once we have covariant expression for energy-momentum of an object,
the rest of the story is straightforward. The volume change we
neglected in the previous calculation can also be treated in a
covariant expression $\mathd \bar{V}({\bar u})$, then the entropy
change of the object is
\begin{equation}
  \mathd S = \beta_{\mu} \mathd P^{\mu} ( \bar{u} ) -  \beta_{\mu} \pi
\mathd  V^{\mu} ( \bar{u} )~,
\end{equation}
where $\beta_\mu = \beta u_{0\mu}$ with $\bar u_0$. This expression is
a revised version of Equation (25) in van Kampen's paper\footnote{The
treatment of the volume is somewhat different in the paper by Israel;
he defined the volume as a fixed region in the three dimensional
space.  Consequently there is no volume change in his theory, and the
effect of compression/expansion is treated through the particle number
flux ($\delta n_{\mu}$ in his paper).}.

With (8) and (9) we obtain
\begin{equation}
  \mathd S = \frac{\beta u_{0 \nu} V_0 u^{\mu}}{u_{0 \lambda} u^{\lambda}}
  \mathd T^{\nu}_{\mu} - \beta u_{0 \mu} \pi \mathd V^{\mu}~.
\end{equation}
The energy-momentum tensor $T$ in equilibrium is expressed as
\begin{equation}
  T_{\mu}^{\nu} = ( \pi + \varepsilon ) u_{0\mu} u_0^{\nu} 
+ \pi \eta_{\mu}^{\nu}~,
\end{equation}
where $\varepsilon$ is the energy density measured in the
comoving frame. Substituting the above expression and (7) into (10) we can
write
\begin{equation}
  \mathd S = \beta V_0 \mathd \varepsilon - \beta \pi \mathd V_0~.
\end{equation}
We understand from the above expression that the entropy of the object
does not depend on $\bar{u}$.  

Now we have the clearly covariant definition of the entropy, other
thermodynamical quantities can be derived covariantly using it.

\end{document}